# Grain-Size Dependence of Thermal and Electrical Properties in (Cr,Mo,Ta,V,W)$C_{1-\delta}$ High-Entropy Carbide Ceramics

Ali Sarikhani[1,*], Gregory E. Hilmas[2], David W. Lipke[2], Douglas E. Wolfe[3], Stefano Curtarolo[4], Shen J. Dillon[5], William G. Fahrenholtz[2,*]

[1] *Missouri University of Science and Technology, Materials Research Center, Rolla, MO, USA*
[2] *Missouri University of Science and Technology, Materials Science and Engineering, Rolla, MO, USA*
[3] *Pennsylvania State University, Materials Science and Engineering, University Park, PA, USA*
[4] *Duke University, Mechanical Engineering and Materials Science, Center for Extreme Materials, Durham, NC, USA*
[5] *University of California, Irvine, Materials Science and Engineering, Irvine, CA, USA*

***ORCID:***

***Sarikhani****: 0000-0001-8121-0867*
***Hilmas****: 0000-0002-1611-5457*
***Lipke****: 0000-0002-4557-6690*
***Wolfe****: 0009-0004-5090-0290*
***Curtarolo****: 0000-0003-0570-8238*
***Dillon****: 0000-0002-6192-4026*
***Fahrenholtz****: 0000-0002-8497-0092*

## Abstract

Thermal and electrical transport in high-entropy carbides can be influenced by grain-boundary scattering, but the relevant microstructural regime remains uncertain. Here, near-fully dense (Cr,Mo,Ta,V,W)$C_{1-\delta}$ ceramics were densified by spark plasma sintering at 1750-1950 °C, producing average grain sizes from 9.3 ± 4.9 to 28.8 ± 15.3 μm while retaining the rock-salt phase. Heat capacities were measured experimentally and estimated by the Neumann-Kopp law with the latter used to calculate thermal conductivity. Thermal conductivity values were in the range of 6.8-7.1 W $m^{-1}$ $K^{-1}$ at room temperature, but showed no systematic dependence on grain size within the investigated range. Room-temperature electrical resistivity values were in the range of 151.9 to 177.0 μΩ cm and likewise showed no grain-size dependence. The electron contribution to room-temperature thermal conductivity was estimated to be approximately 58-68% of total thermal conductivity as estimated using the Wiedemann-Franz law and the theoretical Lorenz number. The results indicated that processing-induced grain coarsening did not produce clearly resolvable transport-property changes under the conditions investigated leading to the conclusion that intrinsic disorder and defect chemistry may exert stronger influences on properties.

**Keywords:** High-entropy carbide; spark plasma sintering; grain size; thermal diffusivity; thermal conductivity; electrical resistivity.

* Corresponding authors: A. Sarikhani (as5kw@mst.edu), W. G. Fahrenholtz (billf@mst.edu)

## Introduction

High-entropy carbides (HECs) are chemically complex refractory ceramics in which multiple transition metals occupy a common metal sublattice within a single rock-salt structure [1-4]. Their multi-principal-element chemistry enables compositional design while preserving the intrinsically high melting temperatures, hardness, and chemical stability of refractory carbides at elevated temperatures, motivating interest in HECs for extreme-environment thermal and structural applications [1-6]. In recent years, substantial progress has been made in producing dense single-phase HEC ceramics using powders synthesized by carbothermal reduction and densified by spark plasma sintering (SPS) [7-10], as well as in developing computational methods for their discovery [1,11]. These studies have established baseline microstructural and mechanical property trends across a variety of HEC compositions.

Beyond phase formation and densification, understanding thermal and electrical transport in HECs is increasingly important because transport properties influence heat transfer, thermal gradients, thermal-shock response, and electrothermal coupling in high-temperature environments. Several studies have reported thermal conductivity and electrical resistivity in dense HEC ceramics and related high-entropy transition-metal carbides, emphasizing the roles of chemical disorder, defect chemistry, and electronic contributions to heat transport [12,13]. More generally, transport in refractory carbides reflects a balance between electron conduction and lattice (phonon) transport, both of which can be affected by microstructural features such as grain boundaries, porosity, and global and local compositional heterogeneity [12-14].

Grain size is a relevant microstructural parameter because grain boundaries can scatter phonons and charge carriers in electrically conducting ceramics, thereby altering thermal conductivity and electrical resistivity. Models for polycrystalline thermal transport frequently include grain-boundary scattering and/or resistance terms. Both predictive and experimental studies show that reducing grain size can reduce thermal conductivity through increased boundary scattering [15,16]. Even when intrinsic point-defect or alloy scattering is strong, grain-boundary density can remain a significant tuning variable depending on the characteristic mean free paths of heat carriers [16-18]. Therefore, comparing carrier mean free paths with the grain-size range provides a useful criterion for evaluating whether grain-boundary scattering should measurably influence thermal and electrical transport in the present specimens.

(Cr,Mo,Ta,V,W)$C_{1-\delta}$ was selected as a model HEC for the present study because it forms a single-phase rock salt structure. It also has a relatively high valence electron concentration, which has been associated with increased metallic bonding character while retaining the high hardness characteristic of refractory carbides [19-22]. The thermophysical properties and grain-growth behavior of (Cr,Mo,Ta,V,W)$C_{1-\delta}$ have been investigated in our previous studies [9,10], providing an established basis for examining the relationship between microstructure and thermal and electrical transport properties. Although the thermophysical properties of this composition have been reported previously [10], the influence of grain size on these transport properties has not been examined.

For HEC ceramics, assessing grain-size effects on properties is challenging because grain growth often occurs concurrently with changes in density, phase constitution, or defect state, making it difficult to attribute property changes solely to grain size. SPS processing provides a useful route for investigating this relationship because dense, nominally single-phase ceramics can be produced at different temperatures to generate systematic grain-size variations [7,8,23]. Here, this approach was used to produce a grain-size series from 9.3 to 28.8 μm while maintaining near-full density and a stable rock-salt phase. This controlled processing series enables evaluation of whether thermal and electrical transport vary systematically with grain size within the investigated coarse-grained regime. Accordingly, the objective of this study was to evaluate the influence of grain size on the thermal and electrical transport properties of (Cr,Mo,Ta,V,W)$C_{1-\delta}$ over the experimentally accessible range of 9.3-28.8 µm. The present work is intended as an initial assessment of grain-size-dependent transport properties while properties at elevated temperatures will be addressed separately.

## Methods and Experiments

### *Carbothermal Reduction and Densification*

The following commercial powders were used as starting materials: carbon black (C120; Cabot, Alpharetta, GA); vanadium (V) oxide ($V_2O_5$, 99.6%, -10 mesh; Alfa Aesar); chromium (III) oxide ($Cr_2O_3$, 99.5%, 0.7 µm; Elementis, Corpus Christi, TX); molybdenum (VI) oxide ($MoO_3$, 99.9%, 6 µm; US Research Nanomaterials, Houston, TX); tantalum (V) oxide ($Ta_2O_5$, 99.8%, 1-5 µm; Atlantic Equipment Engineers, Upper Saddle River, NJ); and tungsten (VI) oxide ($WO_3$, 99.9%, ~80 nm; Inframat Advanced Materials, Manchester, CT). All specimens were made from an identical batch with starting powders of 1.5608 g $Cr_2O_3$, 2.9561 g $MoO_3$, 4.5377 g $Ta_2O_5$, 1.8676 g $V_2O_5$, 4.7613 g $WO_3$, and 3.9927 g carbon black, as described in our previous paper [10]. Powders were batched for a nominally equiatomic mixture of metals and then blended by high-energy ball milling (SPEX 8000D MIXER/MILL, Glen Mills Inc.) in a WC vial using WC media, with a WC media-to-powder mass ratio of 15:2. Two identical ~20 g batches were prepared, passed through a 60-mesh sieve, and compacted into pellets using steel die sets.

Carbothermal reduction was performed by reacting the pellets at 1610 °C for 3 h under a mild vacuum (~13.3 Pa) in a graphite-element furnace (HP50-7010 G, Thermal Technology). Reaction was carried out using a nominally carbon-deficient batching strategy selected to produce single-phase ceramics without residual oxides or excess carbon inclusions. The resulting composition is described using the notation $(Cr,Mo,Ta,V,W)C_{1-\delta}$, where the carbon content was decreased by 7.5 wt% with respect to the stoichiometric amount, corresponding to a nominal Vacancy fraction of 0.13. Reacted pellets were crushed and passed through a 100-mesh sieve. Densification was accomplished by SPS at five temperatures from 1750 to 1950 °C, producing a series of materials with different grain sizes. The SPS cycle began under vacuum (<6 Pa) with heating to 1600 °C at 100 °C/min under an initial uniaxial pressure of 15 MPa, followed by a 5 min dwell. Pressure was then increased to 50 MPa at 35 MPa/min, and temperature was raised to the final sintering value (1750-1950 °C) at 100 °C/min under constant pressure. After a 10-min dwell at the peak temperature, pressure was reduced to 25 MPa at 25 MPa/min while cooling to 1200 °C at 50 °C/min. Lastly, the temperature was reduced to room temperature at the maximum natural cooling rate of the SPS system, while the pressure was lowered to 5 MPa at 20 MPa/min.

Ram displacement was recorded throughout the SPS cycle and used to evaluate densification behavior. Here, ram displacement data were corrected for thermal expansion of the tooling assembly. The thermal expansion correction factor was determined from the slope of ram position versus pyrometer temperature during the 25 MPa stage using a linear fit over a selected temperature interval. The corrected displacement at each temperature was obtained by subtracting the thermal expansion contribution of the die setup from the measured ram displacement. Net sample shrinkage, $\Delta L$, was calculated as the difference between the corrected displacement at the end of the sintering cycle and that at a given time. Assuming uniaxial shrinkage along the pressing direction, the instantaneous relative density was estimated from the final density and the normalized change in sample thickness (using an initial constant thickness), yielding densification curves as a function of temperature.

*Microstructure, Phase Characterization, and Density*

High-contrast microstructures for grain-size analysis were obtained by scanning electron microscopy (SEM; Axia ChemiSEM, Thermo Fisher Scientific Inc.). Grain size was quantified by manually tracing grain boundaries using the GIMP image manipulation software (The GNU Image Manipulation Program Team [24]), and measuring Feret diameters using image analysis software (ImageJ, National Institutes of Health, Bethesda, MD). For each sintering condition, 200-500 grains were analyzed and uncertainty was reported as standard deviations. Representative grain-size, lattice-parameter, and density values for the five sintering conditions are summarized in Table 1. Additional SEM imaging and chemical analysis by energy-dispersive spectroscopy (EDS), including elemental mapping and point quantification, were performed using a second SEM system (PIONEER, Raith 150 eLine Plus).

Phase constitution was evaluated by room-temperature X-ray diffraction (XRD; X'Pert MPD, Philips). For lattice-parameter determination, sintered pellets were crushed and passed through a 100 mesh sieve, and the resulting powders were mixed with α-$Al_2O_3$ as an internal standard ($Al_2O_3$, 99.8%, D50 ~0.46 µm, alpha alumina; Almatis, Leetsdale, PA). Lattice parameters were obtained by Rietveld refinement, and the associated uncertainty was estimated from the calibration/peak-shift correction using the internal standard. Theoretical density was calculated assuming a rock salt carbide structure using the refined lattice parameter together with the average metal-site composition obtained from EDS. The uncertainty in theoretical density was estimated from the standard error of the mean of the EDS measurements, which provided the dominant contribution to the uncertainty in these calculations. Bulk density was measured by the Archimedes method, and its standard error of the mean was determined by propagating the measurement uncertainties of the masses used in the density calculation. Open porosity was also determined from the Archimedes measurements using the corresponding dry, saturated, and immersed specimen masses. Relative density was computed as the ratio of bulk to theoretical density, with uncertainty obtained by standard error propagation from the bulk and theoretical density uncertainties. Based on the measured mass change of the WC milling media before and after milling, each specimen incorporated an average of ~0.03 g WC contamination; this contribution was accounted for in the corrected theoretical density and relative density values reported here. The 2-3% uncertainty in relative density is primarily associated with the uncertainty in the elemental compositions measured by EDS that were used in the theoretical-density calculations. Therefore, nominal values slightly exceeding 100% are interpreted as near-full densification within experimental uncertainty rather than bulk densities that exceeded the predicted theoretical density values. The resulting lattice parameters and density values for the five sintering conditions are summarized in Table 1.

| Sample | Sintering temperature (°C) | Lattice parameter (Å) | Avg. grain size (μm) | Theoretical density ($g/cm^3$) | Bulk density ($g/cm^3$) | Relative density (%) | Open porosity (%) |
|---|---|---|---|---|---|---|---|
| **S1750** | 1750 | 4.2767 ± 0.00015 | 9.3 ± 4.9 | 10.5 ± 0.3 | 10.51 ± 0.03 | 100 ± 3 | 0.12 |
| **S1800** | 1800 | 4.2773 ± 0.00009 | 13.3 ± 8.1 | 10.4 ± 0.2 | 10.46 ± 0.04 | 100 ± 2 | 0.38 |
| **S1850** | 1850 | 4.2775 ± 0.00001 | 18.7 ± 11.3 | 10.3 ± 0.2 | 10.53 ± 0.17 | 102 ± 3 | 1.42 |
| **S1900** | 1900 | 4.2793 ± 0.00010 | 21.5 ± 14.2 | 10.6 ± 0.3 | 10.66 ± 0.03 | 100 ± 3 | 0.16 |
| **S1950** | 1950 | 4.2806 ± 0.00003 | 28.8 ± 15.3 | 10.8 ± 0.2 | 10.55 ± 0.01 | 98 ± 2 | 0.11 |

**Table 1.** Specimen descriptions including sintering temperature, lattice parameter, average grain size (Feret diameter), theoretical density, bulk density, relative density, and open porosity of $(Cr,Mo,Ta,V,W)C_{1-\delta}$ ceramics used to evaluate the grain-size dependence of thermal and electrical transport properties.

*Thermal and Electrical Characterization*

The HEC molar heat capacity was calculated using the Neumann-Kopp rule. The values for the constituent monocarbides that were thermodynamically stable (i.e., TaC) were taken from the NIST database [25]. For the monocarbides that were not stable (i.e., CrC, MoC, VC, and WC with rock salt structures), the quantities were approximated using the Neumann-Kopp rule from the molar heat capacities of the metal and carbon from the same database. This approximation was used because experimental heat-capacity data for the corresponding fictive rock salt monocarbides are not available, although solid-solution carbide data, such as (Ti,W)C, could provide a more refined estimate where sufficient thermodynamic data exist. To interpolate the estimated heat capacity value at the temperatures that thermal diffusivity was measured, a quadratic fit of heat capacity datapoints was used for the calculation of thermal conductivity; i.e., from room temperature to 200 °C in 25 °C increments. The predicted molar heat capacity values ($C_{pm}$, in J/mol K) were then converted to heat capacity at constant pressure on a mass basis ($C_p$, in J/kg K) to obtain SI units. Heat capacities were also measured by differential scanning calorimetry (DSC) and laser flash analysis (LFA) to assess the reliability of the Neumann-Kopp approximation. The values are compared in the Supplementary Information in Figure S1.

Thermal diffusivity was measured using the laser-flash method (Flashline L-S2, Anter Corporation) following ASTM E1461. Measurements were made in 25 °C intervals from room temperature to approximately 200 °C. Specimens were prepared as rounded-corner squares (~12 mm) and were coated with a thin graphite layer to improve energy absorption during testing. Thermal conductivity was then calculated from the measured thermal diffusivity using the corresponding Archimedes density and Neumann-Kopp-predicted heat capacity. The predicted heat capacity was combined with the laser-flash thermal diffusivity to determine the total thermal conductivity as a function of temperature. The same $C_p$ values were used to calculate thermal conductivities for all specimens to maintain a consistent basis for comparison. For each specimen, thermal conductivity was calculated using the standard relation:

$$\kappa(T) = D(T)\rho C_p(T) \quad [W/(m\,K)] \tag{1}$$

where $\kappa$, $D(T)$, $\rho$, and $C_p(T)$ are the total thermal conductivity, thermal diffusivity, bulk density, and specific heat capacity, respectively. Because measurements were limited to T<200 °C, bulk density was assumed to remain constant over this temperature range.

At each temperature, three consecutive laser-flash measurements were performed on the same physical specimen. These repeated measurements were treated as replicates and were used to characterize measurement repeatability. Thermal diffusivity values are reported as the mean ± standard deviation of the three measurements. Because only one physical specimen was examined for each sintering condition, the technical replicates were not used as independent observations in between-condition statistical tests. As an exploratory assessment of grain-size dependence, linear regression and Spearman rank-correlation analyses were performed at each common measurement temperature using the mean thermal diffusivity of the five physical specimens and their corresponding average grain sizes.

Room-temperature electrical resistivity was measured using the Van der Pauw method (ASTM F76) on square specimens with four edge contacts. A switch box was used to cycle current and voltage leads through multiple configurations to obtain geometry-independent resistivity values. To improve measurement reliability, each specimen was tested at applied currents of 0.2, 0.3, and 0.4 A. The Wiedemann-Franz relation (Eq. 2) was used to estimate the electron contribution to thermal conductivity ($\kappa_e$) at a certain temperature from the measured electrical resistivity ($\rho_e$), using a room-temperature Lorenz number representative of transition-metal carbides ($L = 2.44 \times 10^{-8}\ W\ \Omega/K^2$) [12,13]. The resulting electron contribution was compared with the total thermal conductivity to estimate the phonon contribution to thermal conductivity:

$$\kappa_e = \frac{LT}{\rho_e} \quad [W/(m\ K)] \qquad (2).$$

The total thermal conductivity ($\kappa$) was assumed to consist of electron ($\kappa_e$) and phonon ($\kappa_{ph}$) contributions. The theoretical Lorenz number was used to estimate $\kappa_e$, although defects, impurities, and grain boundaries likely cause deviations from ideal behavior. The error bars for $\kappa_{tot}$ were taken from the propagated uncertainty in the thermal conductivity measurements, as described above. The uncertainties in $\kappa_e$ were calculated by error propagation from the electrical resistivity measurements. The uncertainty in the phonon contribution, $\kappa_{ph}$, was then estimated by error propagation from the uncertainties in $\kappa_{tot}$ and $\kappa_e$, i.e., by adding the $\kappa_{tot}$ and $\kappa_e$ uncertainties.

## Results and Discussion

### *Phase Analysis*

Figure 1 shows the XRD patterns of the specimens sintered from 1750 to 1950 °C over a (2θ) range of ~20-140°. All specimens exhibit the same set of peaks corresponding to the FCC rock salt structure, indicating that the same phase was present in all specimens. The additional peaks marked in the patterns correspond to the α-$Al_2O_3$ standard included during the XRD measurements and were not due to any secondary phases in the HEC specimens. No additional specimen-derived peaks were evident within the detection limit of the measurements. These results indicated that the changes in SPS temperature did not change the phase constitution of the HEC specimens.

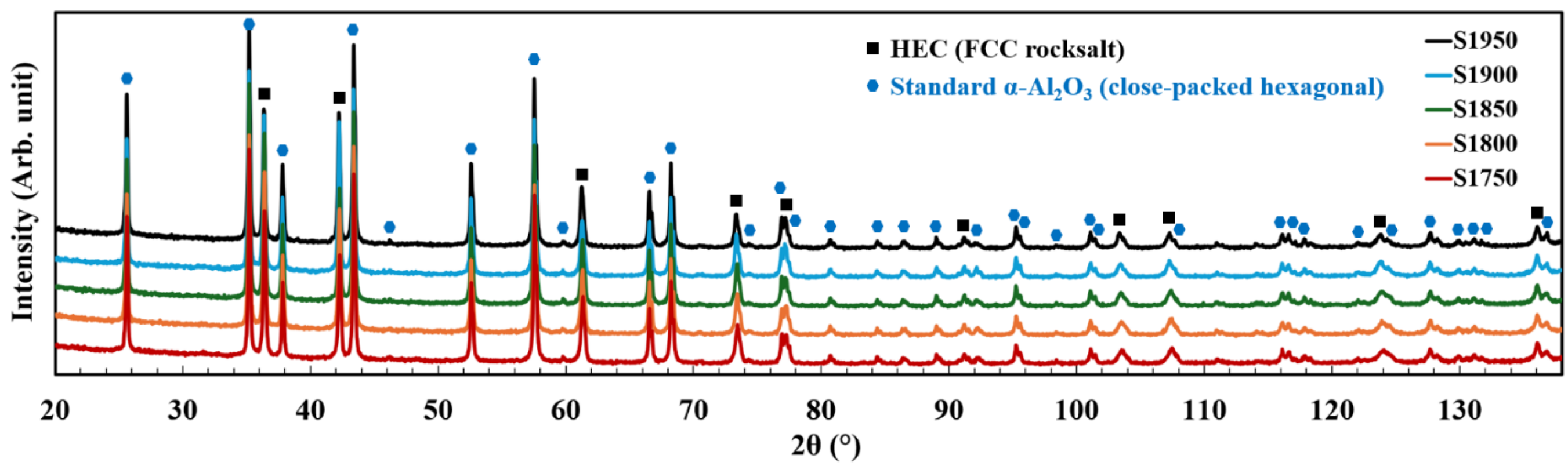

**Figure 1.** X-ray diffraction patterns of (Cr,Mo,Ta,V,W)$C_{1-\delta}$ specimens sintered by SPS at 1750-1950 °C. Peaks corresponding to the FCC rock-salt HEC phase and the α-$Al_2O_3$ standard are identified by black squares and blue hexagonal symbols, respectively.

*Microstructure and elemental homogeneity*

The SPS hold-temperatures ranging from 1750°C to 1950 °C produced near-fully dense, single-phase ceramics with varying grain sizes. Grain-size analysis (Table 1) showed a clear progression from 9.3 ± 4.9 µm at 1750 °C to 28.8 ± 15.3 µm at 1950 °C. The S1950 specimen was produced from a separate powder-processing and SPS batch from the nominally equivalent 1950 °C specimen reported previously in Ref. [10], which may account for the difference in measured grain size between the two studies. Representative SEM micrographs for specimens S1750, S1800, S1850, S1900, and S1950 are shown in Figure 2a-e, respectively. As shown in Figure 2, the SEM micrographs exhibited a systematic coarsening of the microstructure with increasing sintering temperature. The micrographs also showed that the ceramics were fully dense with no obvious residual porosity at the length scales of observation. Although the relative density reported for S1950 was slightly lower (98 ± 2%), this value remained close to full density and no corresponding residual porosity was evident in the SEM observations. The average grain size increased approximately linearly with densification temperature, providing a well-defined microstructural control parameter for evaluating property trends. Because density and phase constitution remained essentially constant across the series, the property comparisons discussed below are interpreted primarily in terms of microstructural length scale (grain size) and intrinsic disorder/defect scattering, rather than porosity-driven extrinsic effects.

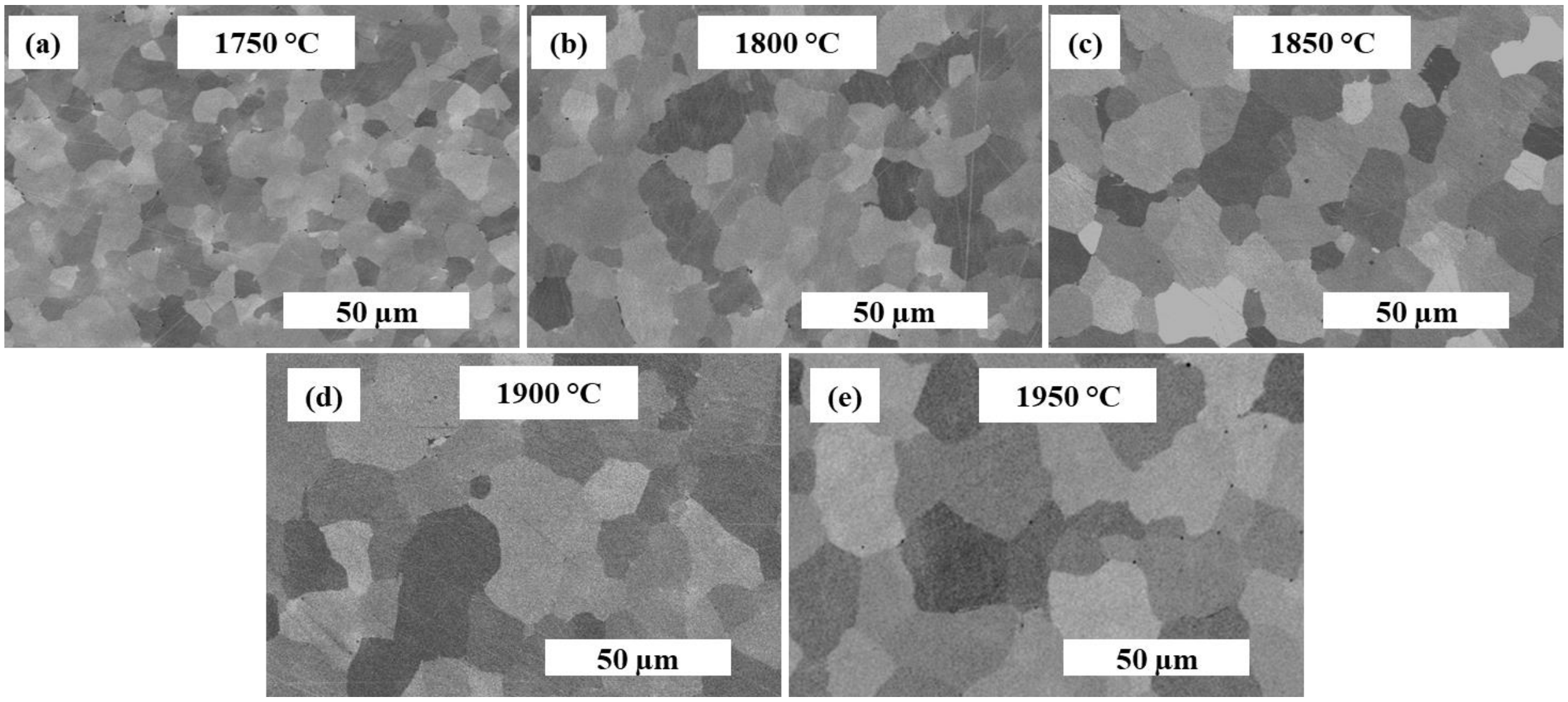

**Figure 2.** Representative high-contrast SEM micrographs used for grain-boundary tracing and Feret diameter measurements for specimens sintered at 1750-1950 °C (a-e), respectively.

Our previous study confirmed that the elemental distributions were homogeneous for all densification temperatures [9]. A combination of backscattered-electron images and corresponding EDS elemental maps showed that Cr, Mo, Ta, V, and W appeared to be uniformly distributed throughout the mapped regions for all processing conditions, with no obvious large-scale elemental segregation detected at the spatial resolution of the EDS measurements. The elemental distributions also remained qualitatively similar for all sintering temperatures. Together with the XRD results, these observations indicate that all of the specimens retained the same overall phase constitution and comparable elemental homogeneity. Note that the previous study reported C and O maps for completeness, but the results were interpreted qualitatively (i.e., they were not used to quantify carbon or oxygen contents of the ceramics) due to the limited accuracy of conventional EDS for light elements.

*Heat capacity*

Heat capacity was predicted using the Neumann-Kopp rule (Figure 3). All specimens had the same nominal composition, which enabled the use of the same $C_p(T)$ relation for the full grain-size series. To assess the validity of the estimated heat capacities, experimental $C_p$ values were measured by DSC and laser flash analysis and compared with the Neumann-Kopp calculations as shown in Figure S1 of the Supplementary Information. The experimentally determined $C_p$ values were used as an independent check on the magnitude and temperature dependence of the Neumann-Kopp estimates rather than as specimen-specific inputs for the thermal-conductivity calculation due to the variability in the experimental measurements. Because the objective of the present study is to compare transport behavior across specimens having the same nominal composition, the same $C_p$ values from the composition-based Neumann-Kopp prediction were used for all specimens to avoid introducing measurement-specific variability into the trends.

Since heat capacity was used directly for calculating thermal conductivity, the predicted $C_p$ contributed to the calculated thermal conductivity for all specimens (see below). A quadratic trendline was then fit (with $R^2$=1) to the resulting heat capacity data to interpolate values at the specific temperatures used for thermal diffusivity measurements, i.e., from room temperature to 200 °C in 25 °C increments. This interpolation provided a consistent $C_p$ value for each thermal-diffusivity measurement temperature and avoided introducing stepwise approximation into the thermal-conductivity calculation. The equation of this quadratic fit is shown in Figure 3.

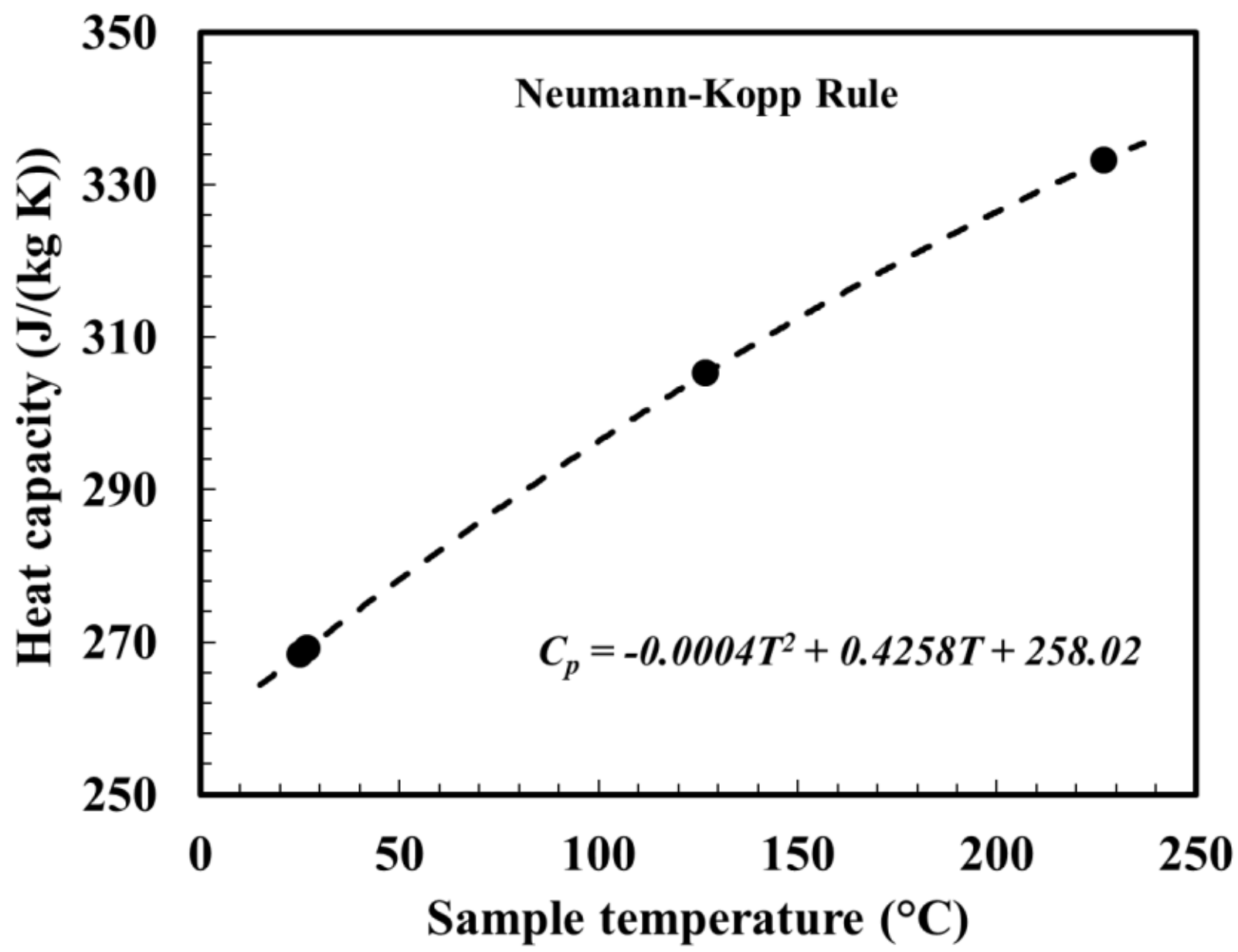

**Figure 3.** Heat capacity predicted by Neumann-Kopp rule as a function of temperature for the nominal $(Cr,Mo,Ta,V,W)C_{1-\delta}$ composition, together with a quadratic fit used for data interpolation.

*Thermal diffusivity*

Thermal diffusivity increased nearly linearly with temperature for all specimens from room temperature to ~200 °C (Figure 4). The diffusivity values for the five sintering conditions were within a narrow band with overlapping error bars. At room temperature, the values were all ~0.025 $cm^2/s$ and increased to ~0.029 $cm^2/s$ at ~200 °C. The error bars represent the standard deviation of three repeated measurements performed on the same physical specimen and quantify measurement repeatability rather than specimen-to-specimen variability. Exploratory regression of the specimen-mean thermal diffusivity against average grain size showed no statistically significant linear association at approximately 50, 75, 125, 150, or 175 °C ($p>0.05$), as represented in Table S1. A statistically significant negative association was detected near 100 °C with $R^2=0.937$ and $p=0.0069$, which was also supported by an exact Spearman rank test ($\rho=-1.0$ and p=0.0167). Because this association was isolated and was not reproduced with statistical significance at adjacent temperatures, the data do not demonstrate a robust grain size dependence across the investigated temperature range. Given the limited number of physical specimens and the absence of independently replicated specimens at each processing condition, these statistical analyses are interpreted as exploratory. Despite this isolated difference, the similar diffusivity values for all of the materials indicated that no consistent systematic grain size dependence was resolved across the investigated temperature range. This behavior suggested that grain size and, therefore, grain boundary density were not the dominant thermal resistance in this temperature range, and may instead reflect a stronger contribution from intrinsic disorder/defect scattering within the bulk lattice. Despite the systematic increase in grain size from S1750 to S1950, no corresponding trend in thermal diffusivity was observed at a given temperature.

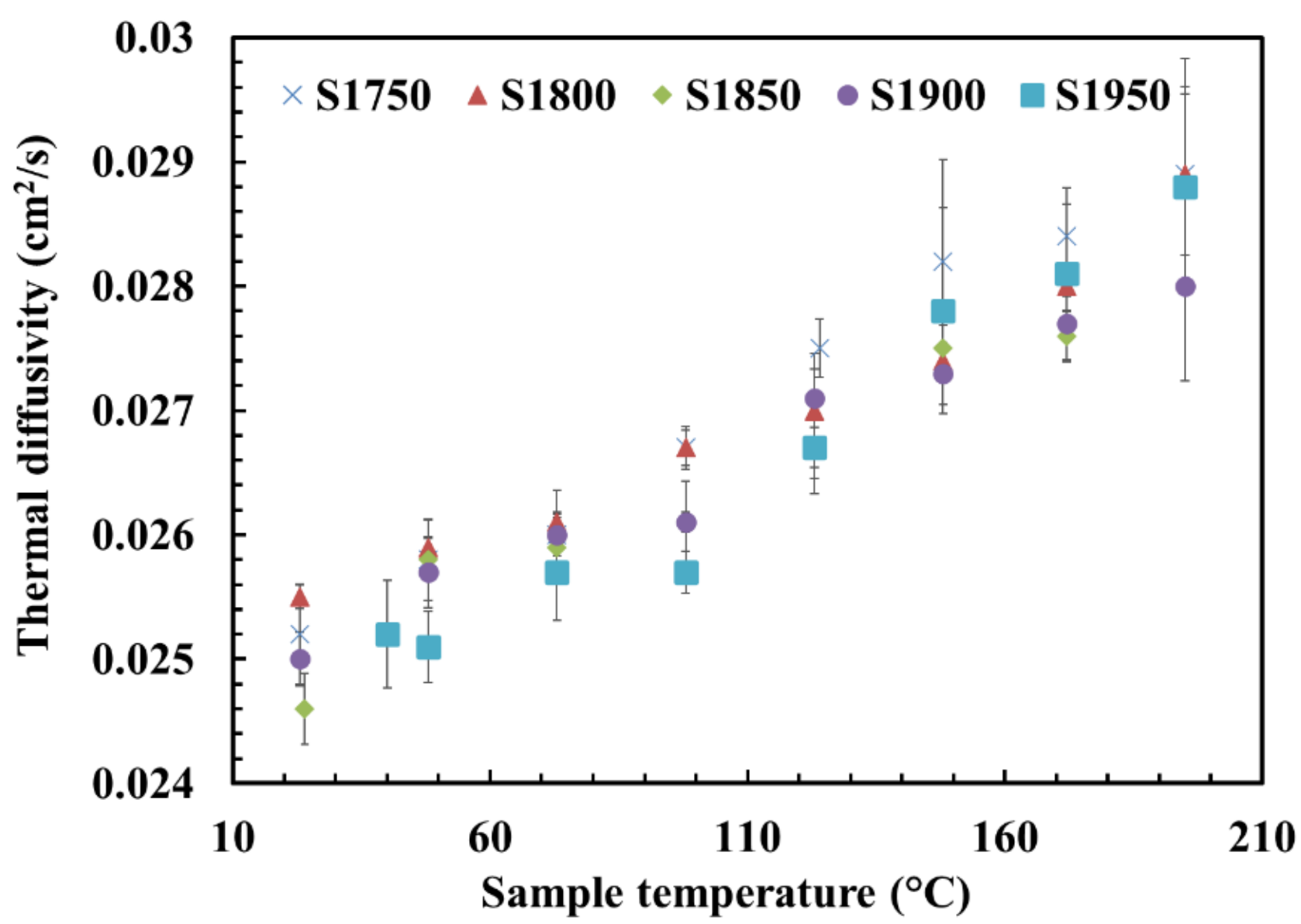

**Figure 4.** Thermal diffusivity from room temperature to ~200 °C for $(Cr,Mo,Ta,V,W)C_{1-\delta}$ ceramics sintered at different temperatures.

*Thermal conductivity and grain-size sensitivity*

Thermal conductivity calculated from measured diffusivity, predicted heat capacity, and measured bulk density increased with temperature for all specimens (Figure 5). The overall temperature dependence was similar across the grain size series, with thermal conductivity increasing from approximately 6.8-7.1 W $m^{-1}$ $K^{-1}$ at room temperature to approximately 9.7-10.1 W $m^{-1}$ $K^{-1}$ near 200 °C. Although modest specimen-to-specimen differences were present, the curves remained closely grouped and did not show any systematic dependence on sintering temperature or grain size. Because thermal diffusivity was nearly the same for all specimens (Figure 4) and the same Neumann-Kopp heat capacity relation was used for the full series (Figure 3), the remaining differences in $\kappa$ are attributed primarily to small differences in bulk density rather than to a strong grain size controlled diffusivity effect. Thus, no clear systematic dependence of thermal conductivity on grain size was resolved within the investigated grain size range.

More broadly, the lack of a clear grain size dependence in $D(T)$ and $\kappa(T)$ suggests that grain boundary scattering was not the dominant factor for heat flow in this system over the measured temperature range. Instead, the transport behavior was consistent with a stronger contribution from intrinsic disorder/defect scattering within the bulk lattice. This interpretation is reasonable for multi-principal-element carbides, where chemical disorder of the metals and vacancy-related carbon site defects are expected to reduce carrier mean free paths (as estimated later) and thereby diminish sensitivity to grain boundary density once a dense, single-phase microstructure is achieved.

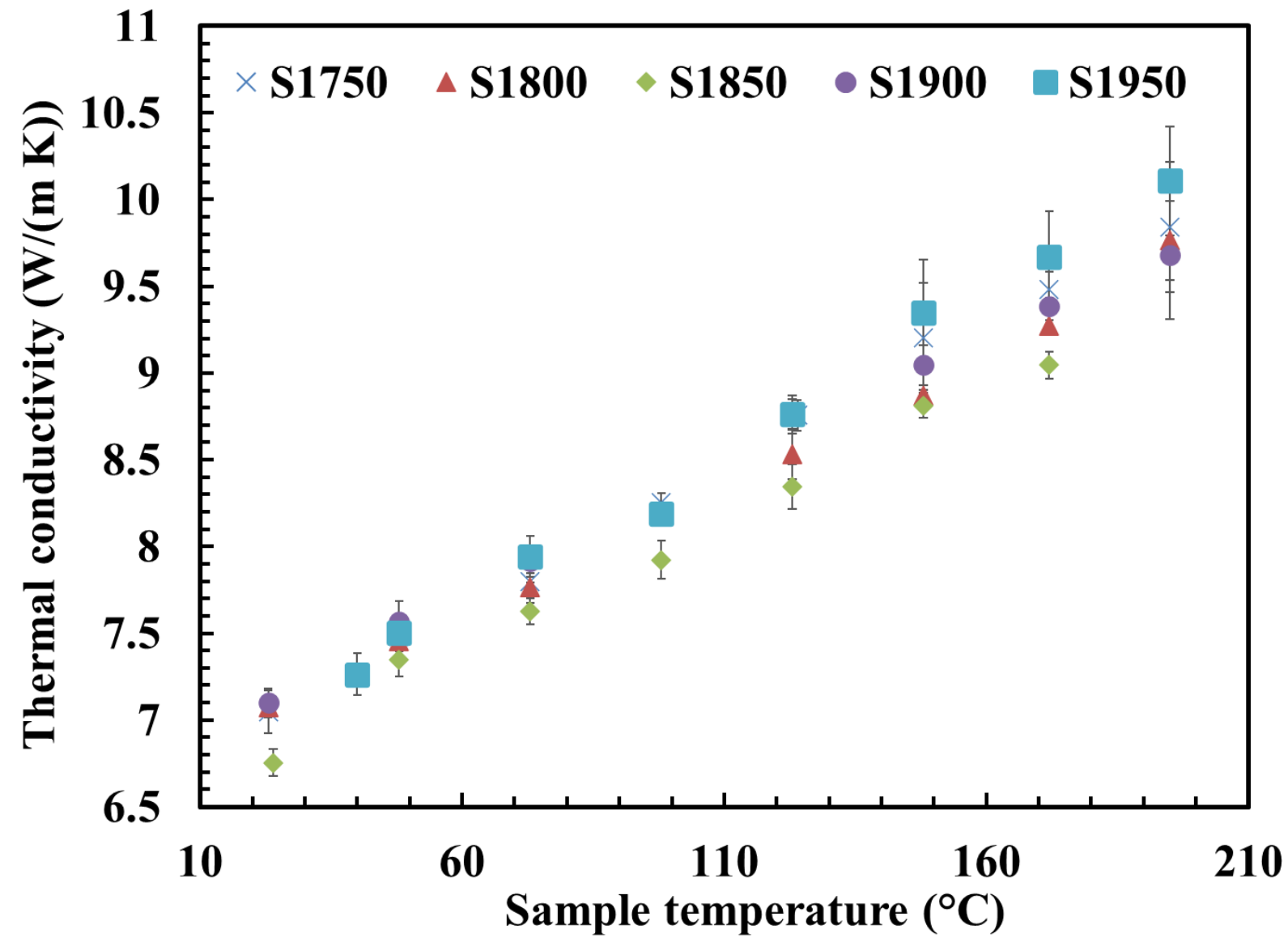

**Figure 5.** Thermal conductivity calculated from measured thermal diffusivity, Neumann-Kopp-predicted heat capacity, and bulk density for (Cr,Mo,Ta,V,W)$C_{1-\delta}$ ceramics sintered at different temperatures.

*Electrical resistivity and electron contribution to thermal conductivity*

Room-temperature electrical resistivity values ranged from 151.9 μΩ cm to 177.0 μΩ cm (Figure 6a). The lowest resistivity was observed for S1750 (151.9 μΩ cm), while S1800 and S1850 were higher (~176-177 μΩ cm), and S1900-S1950 were ~160 μΩ cm. No systematic dependence of electrical resistivity on grain size was observed. This lack of grain size dependence is consistent with the electron mean free path estimated using the Drude/free-electron approximation shown as Equation 3:

$$l_e = \frac{\hbar(3\pi^2)^{1/3}}{e^2\rho_e n^{2/3}} \quad (3),$$

where $l_e$ is the electron mean free path, $\hbar$ is the reduced Planck constant, e is the charge on one electron, $\rho_e$ is the electrical resistivity, and n is the carrier concentration. Using the measured room-temperature resistivity values of ~152-177 μΩ cm and an estimated carrier concentration corresponding to ~1-5 conduction electrons per formula unit gives $l_e$≈0.2-0.6 nm. Even if the carrier concentration is lower, $l_e$ remains in the range of a few nanometers, similar to another report for a high entropy carbide [26]. These values are several orders of magnitude smaller than the measured grain sizes of 9.3-28.8 μm, indicating that electron scattering from grain boundaries is not expected to dominate the room-temperature resistivity in this grain size range. Similarly, phonon mean free paths in chemically complex, vacancy-containing carbides are expected to be limited by intrinsic disorder and point defect scattering, further supporting the lack of grain size dependence on thermal transport properties observed here. Because all of the specimens were nearly fully dense and single-phase, these differences likely reflect specimen-to-specimen variability in defect states and local chemistry (including possible light-element content and short-range inhomogeneity), rather than a direct grain size effect.

The electron contribution to thermal conductivity was estimated using the Wiedemann-Franz relation. The $\kappa_e$ values for all of the materials were in the range of 4 to 5 W m $^{-1}$ K$^{-1}$ (Figure 6b). Like the electrical resistivity, $\kappa_e$ did not exhibit a systematic dependence on grain size. Combining the Wiedemann-Franz estimate with the total thermal conductivity indicated that thermal transport remained electron-dominated across the full grain size range, with $\kappa_e$ contributing about 58-68% of $\kappa_{tot}$, with the assumption that the theoretical Lorenz number was valid for all of the specimens. The remaining contribution, $\kappa_{ph}$, was approximately 2.2-3.0 W m$^{-1}$ K$^{-1}$ and also showed only modest variation with grain size. Thus, neither the electron nor phonon contributions were dependent on grain size within the investigated range.

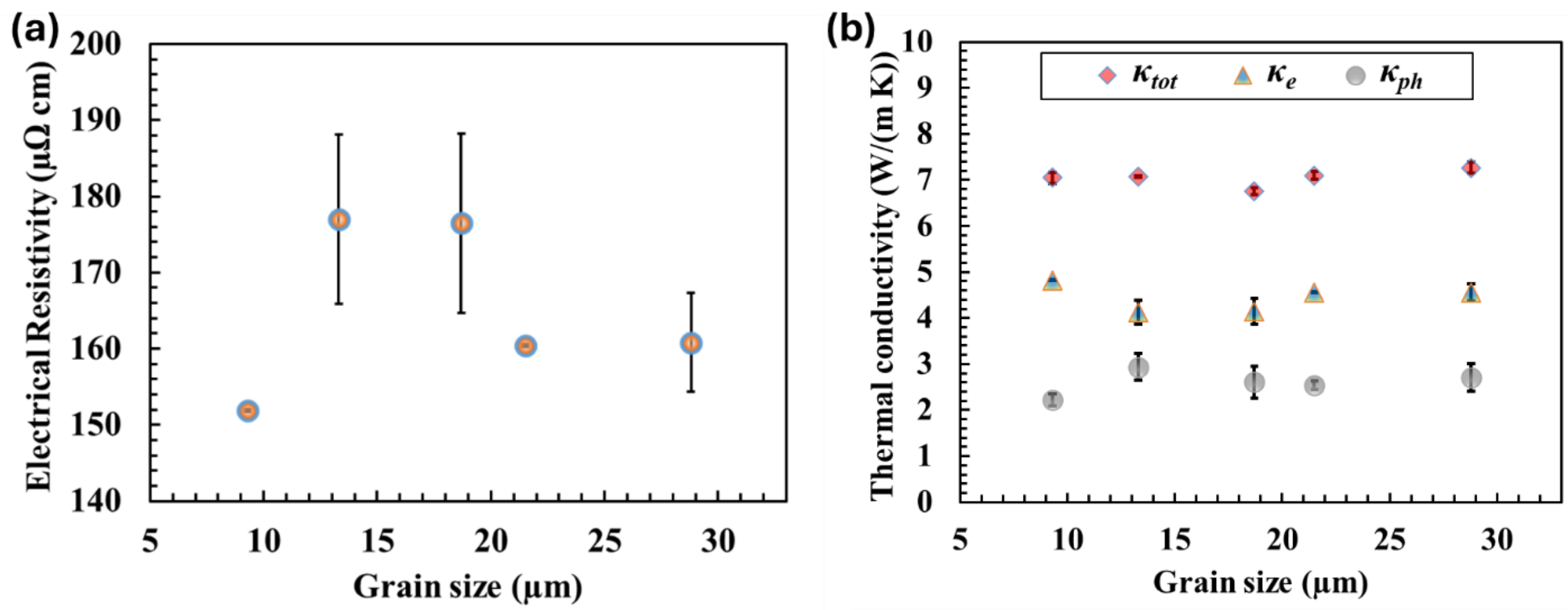

**Figure 6.** (a) Room-temperature electrical resistivity (Van der Pauw) as a function of grain size. (b) Room-temperature total thermal conductivity ($\kappa_{tot}$) as a function of grain size, together with the Wiedemann-Franz electronic contribution ($\kappa_e$) and the estimated phonon contribution ($\kappa_{ph}$).

Although the specimens exhibit the same overall rock salt crystal structure and have comparable distributions of the transition metals, variations in defect chemistry cannot be excluded as a reason for the differences in transport properties. In particular, carbon-vacancy concentration, dissolved oxygen content, and milling-derived WC contamination may influence both electron and phonon transport. Conventional EDS does not provide sufficient accuracy for quantitative determination of carbon or oxygen concentrations in these specimens; therefore, these contributions cannot be independently separated from the influence of grain size. Consequently, the absence of a clear systematic transport trend is interpreted specifically within the investigated grain size range rather than as evidence that grain size does not impact thermal and electrical transport properties in other grain size ranges, particularly when grain sizes become comparable to the mean free paths of the heat and charge carriers.

Across the microstructural series, grain size increased with densification temperature from 9.3 ± 4.9 μm at 1750 °C to 28.8 ± 15.3 μm at 1950 °C, but thermal diffusivity and electrical resistivity showed no significant dependence on sintering temperature. These results indicated that within this near-fully dense, nominally single-phase specimen series, no clear systematic dependence of bulk transport on grain size was resolved. The room-temperature Wiedemann-Franz analysis further showed that the electron contribution to thermal conductivity was significant and approximately constant across conditions, consistent with the estimated sub-nanometer electron mean free paths relative to the micrometer-scale grain sizes. These results support the hypothesis that intrinsic disorder/defect scattering and electron heat transport dominated over grain-boundary-controlled effects in this processing window.

**Conclusions**

Near-fully dense, single-phase $(Cr,Mo,Ta,V,W)C_{1-\delta}$ high-entropy carbide ceramics were produced with grain sizes ranging from 9.3 ± 4.9 μm to 28.8 ± 15.3 μm. Grain size increased systematically with sintering temperature, but thermal diffusivity and thermal conductivity measured between room temperature and 200 °C exhibited only modest sample-to-sample variation with no clear dependence on grain size within the investigated window, as supported by statistical analysis of the thermal diffusivity data. Room-temperature electrical resistivity likewise varied within a narrow range across the series without a systematic grain size trend. The electron contribution to thermal conductivity was estimated to be 60 to 70% and nearly constant across the grain size series, indicating that transport remained electron-dominated under the present conditions. Together, these results indicate that once full density and single-phase stability are achieved, bulk transport in $(Cr,Mo,Ta,V,W)C_{1-\delta}$ is comparatively insensitive to grain size over the microstructural range produced herein. This conclusion is consistent with estimates that the carrier mean free path was on the order of nanometers and more than two orders of magnitude smaller than the grain size. The results suggest that intrinsic disorder/defect scattering may control thermal conductivity in this chemically complex carbide. Overall, this study establishes baseline grain size-property relationships for a dense high entropy carbide and provides guidance for processing-driven microstructural tailoring when property changes are desired.

## Acknowledgements

The authors acknowledge Advanced Materials Characterization Core and the Materials Research Center of the Missouri University of Science and Technology for providing the facilities and resources used in this work. The authors thank Dr. Eric Bohannan from Missouri S&T and Drs. Xiomara Campilongo, Simon Divilov, and Scott Thiel from Duke University for their helpful support and fruitful discussion. This work was part of a Multidisciplinary University Research Initiative project supported by the Office of Naval Research under Award No. N00014-24-1-2768.

## Contributions

**A. Sarikhani** (Conceptualization, Methodology, Visualization, Validation, Writing - original draft, Data curation), **G. E. Hilmas** (Conceptualization, Supervision, Validation, Writing - review & editing), **D. W. Lipke** (Conceptualization, Validation, Writing - review & editing), **D. E. Wolfe** (Conceptualization, Writing - review & editing), **S. Curtarolo** (Conceptualization,

Writing - review & editing), **S. J. Dillon** (Conceptualization, Writing - review & editing), **W. G. Fahrenholtz** (Conceptualization, Methodology, Supervision, Validation, Writing - review & editing).

# *Supplementary Material*

## Grain-Size Dependence of Thermal and Electrical Properties in (Cr,Mo,Ta,V,W)$C_{1-\delta}$ High-Entropy Carbide Ceramics

Ali Sarikhani[1,*], Gregory E. Hilmas[2], David W. Lipke[2], Douglas E. Wolfe[3],
Stefano Curtarolo[4], Shen J. Dillon[5], William G. Fahrenholtz[2,*]

[1] *Missouri University of Science and Technology, Materials Research Center, Rolla, MO, USA*
[2] *Missouri University of Science and Technology, Materials Science and Engineering, Rolla, MO, USA*
[3] *Pennsylvania State University, Materials Science and Engineering, University Park, PA, USA*
[4] *Duke University, Mechanical Engineering and Materials Science, Center for Extreme Materials, Durham, NC, USA*
[5] *University of California, Irvine, Materials Science and Engineering, Irvine, CA, USA*

***ORCID:***

***Sarikhani**: 0000-0001-8121-0867*
***Hilmas**: 0000-0002-1611-5457*
***Lipke**: 0000-0002-4557-6690*
***Wolfe**: 0009-0004-5090-0290*
***Curtarolo**: 0000-0003-0570-8238*
***Dillon**: 0000-0002-6192-4026*
***Fahrenholtz**: 0000-0002-8497-0092*

### *S1. Heat capacity measurements and calculations*

Specific heat capacity, $C_p$, was experimentally evaluated by differential scanning calorimetry (DSC) using a TA Instruments DSC 2010 (V4.4E) with aluminum pans under a nitrogen purge. The measurements were analyzed using the sapphire-reference procedure recommended by TA Instruments, in which an empty-pan baseline and a sapphire standard of known heat capacity are measured under equivalent conditions, and the calorimetric difference between the sample and baseline is used to calculate $C_p$. Figure S1 compares the resulting DSC measurements with heat capacities obtained from the LFA measurements with respect to a thermographite standard and values calculated using the Neumann-Kopp approach. Although some specimen-to-specimen differences are evident among the experimental methods, the measured values are of comparable magnitude and generally exhibit an increase in heat capacity with increasing temperature.

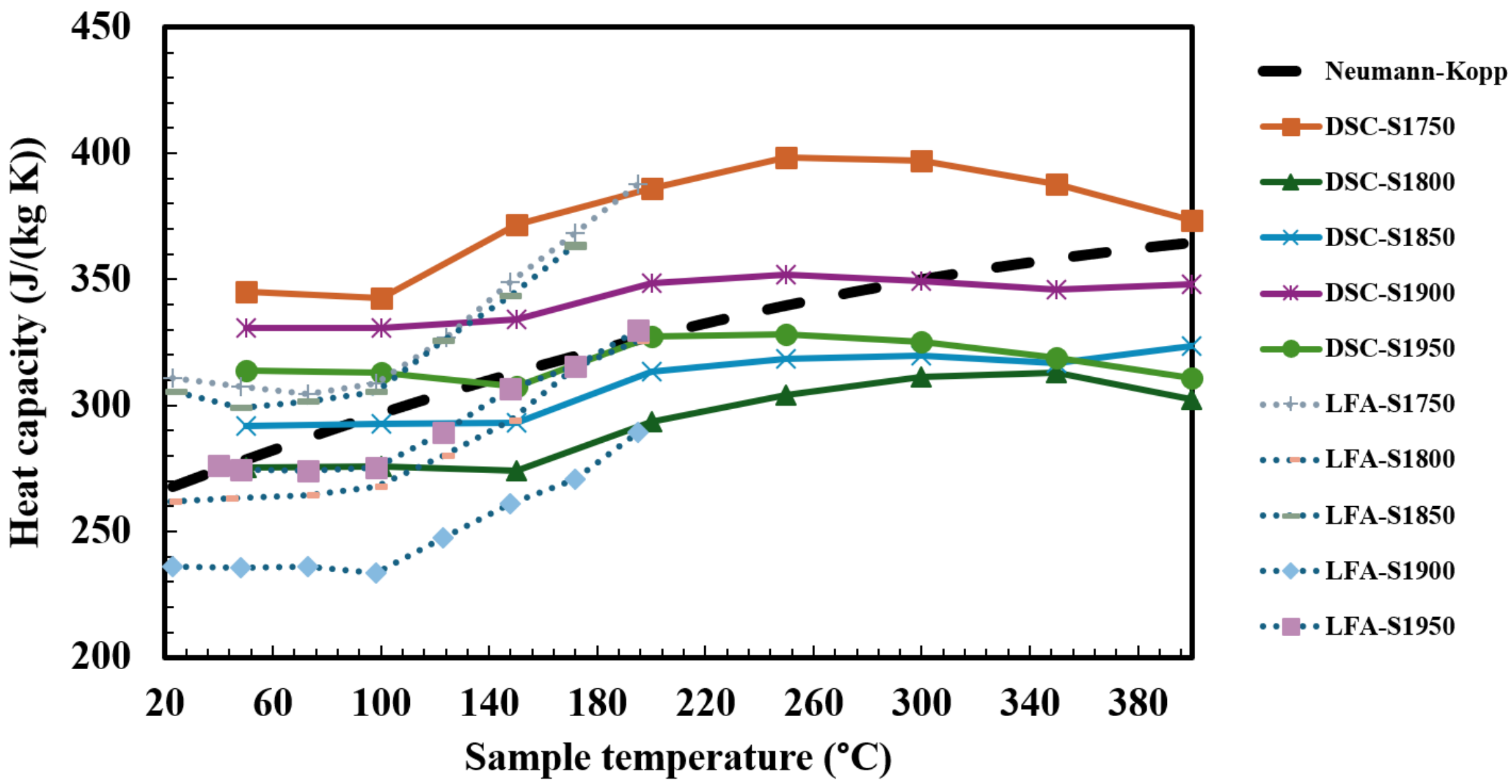

**Figure S1.** Specific heat capacity of (Cr,Mo,Ta,V,W)$C_{1-\delta}$ specimens sintered at 1750-1950 °C. Experimental $C_p$ values obtained by differential scanning calorimetry (DSC) and laser flash analysis (LFA) are compared with values calculated using the Neumann-Kopp approach. The DSC measurements were performed with aluminum pans under nitrogen and analyzed using the sapphire-reference method.

### *S2. Statistical analysis of thermal diffusivity and grain size*

To further assess the relationship between grain size and thermal diffusivity, exploratory statistical analyses were performed at each common measurement temperature using the mean thermal diffusivity of the five physical specimens and their corresponding average grain sizes. Because the repeated laser-flash measurements were multiple flashes on the same specimen, they were treated as technical replicates and used only to characterize measurement repeatability.

Linear regression and Spearman rank-correlation analyses were therefore based on one mean diffusivity value per specimen. As summarized in Table S1, no statistically significant grain-size association was observed at most temperatures, whereas an isolated significant negative relationship was detected near 100 °C. Because this relationship was not reproduced consistently at adjacent temperatures and the number of independently processed specimens was limited, the statistical results are interpreted as exploratory rather than as evidence of a general grain-size dependence.

| Nominal T (°C) | Linear slope ×$10^{-5}$ ($cm^2$ $s^{-1}$ $\mu m^{-1}$) | $R^2$ | Linear-regression p | Spearman ρ | Exact Spearman p |
|---|---|---|---|---|---|
| **50** | -3.31 | 0.676 | 0.087 | -0.70 | 0.233 |
| **75** | -1.75 | 0.580 | 0.135 | -0.80 | 0.133 |
| **100** | **-6.09** | **0.937** | **0.0069** | **-1.00** | **0.0167** |
| **125** | -3.40 | 0.606 | 0.121 | -0.70 | 0.233 |
| **150** | -4.97 | 0.714 | 0.071 | -0.80 | 0.133 |
| **175** | -2.11 | 0.215 | 0.432 | -0.30 | 0.683 |

**Table S1.** Exploratory statistical analysis of the relationship between average grain size and mean thermal diffusivity at each common measurement temperature. Linear-regression slope, coefficient of determination, regression p-value, Spearman rank-correlation coefficient, and exact Spearman p-value are reported. The analysis uses one mean thermal-diffusivity value for each physical specimen; repeated laser-flash measurements on each specimen were treated as technical replicates and were not considered independent observations.